\documentclass[aps,prb,twocolumn]{revtex4}%
\usepackage{amsfonts}
\usepackage{amsmath}
\usepackage{amssymb}
\usepackage{graphicx}
\usepackage{psfrag,color}
\usepackage{hyperref}
\providecommand*{\eu}{\ensuremath{\mathrm{e}}}
\providecommand*{\diff}[1]{\mathrm{d} #1\,}

\begin{document}
\title{Collective diffusion coefficient of proteins with hydrodynamic, electrostatic and adhesive interactions}
\author{Peter Prinsen and Theo Odijk$^{1}$,}
\affiliation{Complex Fluids Theory, Faculty of Applied Sciences, Delft University of
Technology, Delft, the Netherlands}

\begin{abstract}
A theory is presented for $\lambda_{C}$, the coefficient of the
first-order correction in the density of the collective diffusion
coefficient, for protein spheres interacting by electrostatic and
adhesive forces. An extensive numerical analysis of the Stokesian
hydrodynamics of two moving spheres is given so as to gauge the
precise impact of lubrication forces. An effective stickiness is
introduced and a simple formula for $\lambda_{C}$ in terms of this
variable is put forward. A precise though more elaborate
approximation for $\lambda_{C}$ is also developed. These and
numerically exact expressions for $\lambda_{C}$ are compared with
experimental data on lysozyme at pH 4.5 and a range of ionic
strengths between 0.05 M and 2 M.

\end{abstract}

\maketitle

\section*{I. INTRODUCTION}

\footnotetext[1]{Address for correspondence: T. Odijk, P.O. Box 11036, 2301 EA
Leiden, the Netherlands. E-mail: odijktcf@wanadoo.nl}

Fick's first law states that the particle flux is equal to minus
the collective diffusion coefficient times the gradient of the
particle concentration. For colloids or macromolecules in
solution, this collective (also called cooperative or mutual)
diffusion coefficient is often determined experimentally with the
help of dynamic light scattering. If one extrapolates this
coefficient to a vanishing concentration of particles, it reduces to
the single-particle diffusion coefficient since the interactions
between the particles are presumably negligible then. At non-zero volume
fractions, particle interactions, such as those of electrostatic
and hydrodynamic origin, influence the diffusion. At low enough
concentrations, where three- and higher body interactions may be disregarded, the parameter $\lambda_{C}$ characterizes the departure from the single-particle result.

The concentration dependence of the collective diffusion
coefficient of proteins has been studied extensively in
experiments, for example in the case of hemoglobin
\cite{ALP,MIN,HAL,LAG,BER}, bovine serum albumin
\cite{TIN,RAJ,PHI,AND}, $\beta$-lactoglobulin \cite{BON},
ovalbumin \cite{GIB} and lysozyme
\cite{NYS,MIR,MUS,ZHA,SKO,EBE,LEG,PRI,ANN,RET,NAR}. On the
theoretical side, a fair number of papers
\cite{FE0,BRO,OHT,FE1,GEN,PE1,DEN,PE2,GAP} deal with the diffusion
of interacting colloidal particles in solution. Apart from giving
insight into the diffusion as such, the coefficient $\lambda_{C}$
is also important because it could yield information about the
complex pair interaction between protein molecules. Moreover, it
has been argued that $\lambda_{C}$ may be an alternative
parameter useful in diagnosing under what conditions proteins would crystallize
\cite{EBE}.

In Ref.~\onlinecite{PR1} we approximated globular proteins in
water with added monovalent salt by hard spherical particles that
interact through a short-range attraction and a screened
electrostatic repulsion. We appropriately replaced this system by
one of spherical particles with sticky interactions only. At infinite
dilution the effective stickiness is readily determined by
equating the respective second virial coefficients of the two
systems. In the effective stickiness, part of the bare adhesion
is balanced against the electrostatic repulsion.

In the next section, we formulate a theory for the coefficient
$\lambda_{C}$. We first introduce the interaction used previously
to compute protein solution properties \cite{PR1} and give
expressions for the effective stickiness. We then outline the
formal expression for $\lambda_{C}$ due to Felderhof \cite{FE0} in
terms of the pair potential between two protein spheres and a
hydrodynamic mobility function. Although the latter has been
studied often in the past, we present a more extensive numerical
analysis in order to gain more insight into the asymptotics of the
lubrication regime for two moving spheres very close to each other. The
coefficient $\lambda_{C}$ is then computed in three ways: exactly
via numerics and in terms of two convenient approximations. In
section III, we compare these predictions for $\lambda_{C}$ with
experiment. A discussion of the results is given in the last
section.

\section*{II. THEORY}

\subsection*{A. Effective interaction}

We model the globular proteins as spherical particles of radius
$a$ with a total charge $Zq$ per particle that is uniformly
distributed over its surface. Here $q$ is the elementary (proton)
charge. For convenience, we scale all distances by the radius $a$
and all energies by $k_{B}T$ where $k_{B}$ is Boltzmann's constant
and $T$ is the temperature. We approximate the interaction between
two proteins by a steric repulsion plus a short-range attraction
of scaled range $\delta\ll 1$ and constant absolute magnitude
$U_{A}$, and a far-field Debye-H\"uckel potential. The latter describes the Coulomb repulsion that is screened due to the presence of monovalent salt of ionic strength $I$. The effective number $Z_{eff}$ of charges associated with the far field is computed in the Poisson-Boltzmann approximation. See Refs.~\onlinecite{PR1} and \onlinecite{PR2} for further details. The total interaction $U_{T}(x)$ between the two particles
with center-of-mass separation $r$ is thus of the form
\begin{equation}
U_{T}(x)=\left\{
\begin{array}
[c]{ll}%
\infty & \hspace{15pt} 0\leq x<2\\
U_{DH}(x)-U_{A} & \hspace{15pt} 2\leq x<2+\delta\\
U_{DH}(x) & \hspace{15pt} x\geq2+\delta
\end{array}
\right.  , \label{pot}
\end{equation}
\begin{equation}
x\equiv\frac{r}{a}.
\end{equation}
Here, the Debye-H\"uckel interaction is given by
\begin{equation}
U_{DH}(x)=2\xi\frac{\eu^{-\mu(x-2)}}{x} \label{debye}
\end{equation}
where $\xi\equiv\frac{Q}{2a}\left( \frac{Z_{eff}}{1+\mu}\right)
^{2}$ and $\mu\equiv\kappa a$. The Debye length $\kappa^{-1}$ is
defined by $\kappa^{2}\equiv 8\pi QI$ and the Bjerrum length by
$Q\equiv q^{2}/\epsilon k_{B}T$, which equals $0.71$ nm in water
at $298$ K ($\epsilon$ is the permittivity of water);
$\mu=3.28a\sqrt{I}$, if the radius $a$ is given in nm and the
ionic strength $I$ in M. We suppose 1-1 electrolyte has been added
in excess so $I$ is the concentration of added salt only. We have
derived a perturbative expression for the effective charge $qZ_{eff}$ in the Poisson-Boltzmann
approximation \cite{PR1}
\begin{equation}
Z_{eff}=Z-\frac{\mu^{2}}{6}\left(  \frac{Q}{a}\right)^{2} \left(
\frac{Z}{1+\mu }\right)^{3}\eu^{3\mu}E_{1}(3\mu). \label{Zeff}
\end{equation}
Here, $E_{1}(x)$ is the exponential integral defined by
$E_{1}(x)=\int _{x}^{\infty}dt\,t^{-1}\eu^{-t}$. Eq.~(\ref{Zeff})
is numerically consistent with a different form recently proposed
by Aubouy et al. \cite{AUB} which is also valid at large values of
$Z$.

We want to replace the system of particles interacting through the
complicated interaction (\ref{pot}) by a system of particles
interacting through a simpler potential, the adhesive hard sphere
(AHS) potential of Baxter \cite{BAX}
\begin{equation}
U_{AHS}(x)=\left\{
\begin{array}
[c]{ll}%
\infty & \hspace{15pt} 0\leq x<2\\
\ln\frac{12\tau\omega}{2+\omega} & \hspace{15pt} 2\leq x\leq2+\omega\\
0 & \hspace{15pt} x>2+\omega
\end{array}.
\right.\label{ahs}
\end{equation}
Here, $\tau$ is a positive constant which signifies the strength
of the \textit{effective} adhesion and the limit
$\omega\downarrow0$ has to be taken appropriately after formal
integrations. In order to replace the original system by this
simpler system, we have to find the correspondence between the
parameter $\tau$ in the AHS potential and the parameters $\xi$,
$\mu$, $\delta$ and $U_{A}$ in the original interaction
Eq.~(\ref{pot}). In this case, we do this by matching the
respective second virial coefficients, which ensures that the free
energy of the two systems at small concentrations are identical. We
emphasize that in the general case, at arbitrary concentrations,
we have to match the complete free energies of the respective systems \cite{PR1,PR2}; it is then incorrect to focus on the second virials as has often been done in the past.

\subsection*{B. Stickiness parameter}

We already determined the stickiness parameter $\tau$ in a
previous paper \cite{PR1}. Here we reproduce the main results. The
second virial coefficient $B_{2}$ is given by
\begin{equation}
B_{2}= \frac{1}{2}\int_{V}\diff{\mathbf{r}}\left(1-\eu^{-U\left(
\mathbf{r}\right)}\right),
\end{equation}
where $U\left(\mathbf{r}\right)$ is the pair potential scaled by
$k_{B}T$, and $\mathbf{r}$ is the unscaled position vector connecting the
centers of mass of the two particles. For the pair interaction of
Eq.~(\ref{pot}), $B_{2}$ may be expressed by
\begin{equation}
B_{2}=B_{2}^{HS}\left(1+\frac{3}{8}J\right), \label{vir2}
\end{equation}
where we introduce the following integrals
\begin{equation}
J\equiv\int_{2}^{\infty}\diff{x}x^{2}\left(1-\eu^{-U_{T}\left(x\right)
}\right) \equiv J_{1}-\left(  \eu^{U_{A}}-1\right)  J_{2},
\label{vir3}
\end{equation}
\begin{align}
J_{1}&\equiv\int_{2}^{\infty}\diff{x}x^{2}\left(
1-\eu^{-U_{DH}(x)}\right)\nonumber\\ & \simeq\frac{4\left(
\mu+\frac{1}{2}\right) \xi}{\mu^{2}}\left(
1-\frac{\alpha}{2}\xi\right)\label{J1}  ,
\end{align}
\begin{align}
J_{2}&\equiv\int_{2}^{2+\delta}\diff{x}x^{2}\eu^{-U_{DH}(x)}\nonumber \\ & \simeq2\delta\left[
\eu^{-\xi}+\left(  1+\frac{\delta}{2}\right)
^{2}\eu^{-\frac{\xi}{1+\delta/2}\eu^{-\mu\delta}}\right]\label{J2}.
\end{align}
Here, $B_{2}^{HS}=16\pi a^{3}/3$ is the value of $B_{2}$ if the
proteins were solely hard spheres and
$\alpha=\frac{\eu^{-\xi}-(1-\xi)}{\xi^{2}}$. We equate
Eq.~(\ref{vir2}) with the second virial coefficient of the AHS
model
\begin{equation}
B_{2}=B_{2}^{HS}\left(1-\frac{1}{4\tau}\right),\label{B2bax}
\end{equation}
which results in a stickiness parameter $\tau$ given by
\begin{equation}
\tau=-\frac{2}{3J}\label{tb2}.
\end{equation}
From Eqs.~(\ref{pot}) and (\ref{vir3}) we see how part of the original
attraction is compensated by repulsive electrostatics.

\subsection*{C. General expression for $\lambda_{C}$}

For small volume fractions $\phi$ of spherical particles, the collective
diffusion coefficient $D_{C}$ may be written as
\begin{equation}
D_{C}=D_{0}\left(1+\lambda_{C}\phi+O\left(\phi^{2}\right)\right),
\end{equation}
where $D_{0}$ is the diffusion coefficient in the dilute limit.
The linear coefficient $\lambda_{C}$ may be split up into five
contributions \cite{FE0}
\begin{equation}
\lambda_{C}=\lambda_{V}+\lambda_{O}+\lambda_{D}+\lambda_{S}+\lambda_{A}.
\label{lc}
\end{equation}
These terms have been studied for some time
\cite{FE0,BRO,OHT,FE1}: there is a virial correction because a
fluctuation in the osmotic pressure drives diffusion
\begin{equation}
\lambda_{V}=3\int_{0}^{\infty}\diff{x}x^{2}\left(1-\eu^{-U(x)}\right),\label{lv}
\end{equation}
and four terms arising from the mutual friction between two
hydrodynamically interacting spheres. An Oseen contribution
\begin{equation}
\lambda_{O}=3\int_{0}^{\infty}\diff{x}x\left(\eu^{-U(x)}-1\right),\label{lo}
\end{equation}
and a dipolar contribution
\begin{equation}
\lambda_{D}=1,  \label{ld}
\end{equation}
express the long-range hydrodynamic interaction between two
particles 1 and 2 whereas the short-range part of the hydrodynamic
interaction comes into play in the term
\begin{equation}
\lambda_{S}=\int_{2}^{\infty}\diff{x}x^{2}\eu^{-U(x)}\left(A_{12}^{tt}(x)+
2B_{12}^{tt}(x)-\frac{3}{x}\right).\label{ls}
\end{equation}
Finally, the modification of the single-particle mobility is
expressed by
\begin{equation}
\lambda_{A}=\int_{2}^{\infty}\diff{x}x^{2}\eu^{-U(x)}\left(A_{11}^{tt}(x)+
2B_{11}^{tt}(x)\right).\label{la}
\end{equation}
Here, $A_{11}^{tt}(x)$, $A_{12}^{tt}(x)$, $B_{11}^{tt}(x)$ and
$B_{12}^{tt}(x)$ are dimensionless hydrodynamic functions given in
terms of the translational mobility matrix for two spheres
centered at $\vec{R}_{1}$ and $\vec{R}_{2}$
($\vec{r}=\vec{R}_{1}-\vec{R}_{2})$ and acquiring velocities
$\vec{V}_{1}$ and $\vec{V}_{2}$ as a result of the forces
$\vec{F}_{1}$ and $\vec{F}_{2}$ acting on the spheres
\begin{equation}
\vec{V}_{1}=\mu_{11}^{tt}(1,2)\cdot
\vec{F}_{1}+\mu_{12}^{tt}(1,2)\cdot \vec{F}_{2}
\end{equation}
\begin{equation}
\vec{V}_{2}=\mu_{21}^{tt}(1,2)\cdot
\vec{F}_{1}+\mu_{22}^{tt}(1,2)\cdot \vec{F}_{2}\label{v2}
\end{equation}
In the notation of Cichocki and Felderhof \cite{FE2}, we have
\begin{equation}
\mu_{11}^{tt}(1,2)=\frac{1}{6\pi\eta a}\left[\vec{\vec{I}}+
A_{11}^{tt}(r)\frac{\vec{r}\vec{r}}{r^{2}}+
B_{11}^{tt}(r)\left(\vec{\vec{I}}-\frac{\vec{r}\vec{r}}{r^{2}}\right)\right]\label{mu11}
\end{equation}
\begin{equation}
\mu_{12}^{tt}(1,2)=\frac{1}{6\pi\eta a}\left[
A_{12}^{tt}(r)\frac{\vec{r}\vec{r}}{r^{2}}+
B_{12}^{tt}(r)\left(\vec{\vec{I}}-\frac{\vec{r}\vec{r}}{r^{2}}\right)\right],\label{mu12}
\end{equation}
where $\eta$ is the viscosity of the solvent and $\vec{\vec{I}}$
is the unit tensor. The mobility tensors in Eq.~(\ref{v2}) are
given by interchanging the labels in Eqs.~(\ref{mu11}) and
(\ref{mu12}) while taking into account the symmetry relations
\begin{equation}
A_{12}^{tt}(r)=A_{21}^{tt}(r)\text{;}\hspace{1cm}B_{12}^{tt}(r)=B_{21}^{tt}(r).
\end{equation}

Recall that the particles have a hard-core interaction for $x<2$
so $\exp-U(x)$ vanishes for $x<2$. We then sum Eqs.~(\ref{lv})-(\ref{la})
and conveniently rewrite $\lambda_{C}$ as follows
\begin{equation}
\lambda_{C}=c_{0}+c_{1}\int_{2}^{\infty}\diff{x}x^{2}\left(1-\eu^{-U(x)}\right)+
R. \label{aplc1}
\end{equation}
The constant $c_{0}$ equals the value $\lambda_{C}$ would adopt if
the spheres were hard but without any other interaction
\begin{align}
c_{0} & \equiv 3\int_{0}^{2}\diff{x}x^{2}-3\int_{0}^{2}\diff{x}x+1+
\int_{2}^{\infty}\diff{x}x^{2}\left(h(x)-\frac{3}{x}\right)\nonumber \\ & =
3+\int_{2}^{\infty}\diff{x}x^{2}\left(h(x)-\frac{3}{x}\right).
\label{c0}
\end{align}
Here, $h(x)$ is the sum of scalar mobility functions
\begin{equation}
h(x)\equiv A_{11}^{tt}(x)+A_{12}^{tt}(x)+2B_{11}^{tt}(x)+
2B_{12}^{tt}(x).\label{hx}
\end{equation}
The residual term $R$ in Eq.~(\ref{aplc1}) depends on the actual
interaction
\begin{equation}
R\equiv\int_{2}^{\infty}\diff{x}x^{2}\left(\eu^{-U(x)}-1\right)
\left(h(x)-h(2)\right)\label{R}
\end{equation}
though it would vanish if the interaction $U$ were adhesive and
purely of the Baxter type (see Eq.~(\ref{ahs})). The second term
on the right hand side of Eq.~(\ref{aplc1}) is proportional to the
constant
\begin{equation}
c_{1}\equiv 3-h(2) \label{c1}
\end{equation}
and the integral is related to the second virial coefficient
$B_{2}$ by (see Eqs.~(\ref{vir2}) and (\ref{vir3}))
\begin{equation}
\int_{2}^{\infty}\diff{x}x^{2}\left(1-\eu^{-U(x)}\right)=
\frac{8}{3}\left(\frac{B_{2}}{B_{2}^{HS}}-1\right). \label{intv}
\end{equation}
The resulting expression for $\lambda_{C}$ is
\begin{equation}
\lambda_{C}=
c_{0}+\frac{8c_{1}}{3}\left(\frac{B_{2}}{B_{2}^{HS}}-1\right)+ R
\label{aplc2}
\end{equation}
which we can evaluate once we know $h(x)$ given by Eq.~(\ref{hx}).

\subsection*{D. Hydrodynamics}

The function $h(x)$ was discussed by Batchelor \cite{BA1} in his
theory of the diffusion of hard spheres. The sum
$A_{11}^{tt}+A_{12}^{tt}$ pertains to the mobility of a pair of
spheres moving in the direction of their line of centers whereas
$B_{11}^{tt}+B_{12}^{tt}$ is related to their mobility when they
move perpendicular to that line. (Note that in
Ref.~\onlinecite{BA1} $A_{11}\equiv A_{11}^{tt}+1$, $B_{11}\equiv
B_{11}^{tt}+1$, $A_{12}\equiv A_{12}^{tt}$ and $B_{12}\equiv
B_{12}^{tt}$). In the latter case, because the spheres are
couple-free, the spheres must rotate as the pair translates. At
small separations ($x-2\ll 1$), lubrication forces with a
logarithmic singularity $\ln^{-1}(x-2)$ are then expected to
develop on general grounds \cite{BA2}. Goldman et al \cite{GOL}
proposed a form for the singularity which we will test below.

Batchelor \cite{BA1} computed $h(2)=1.312$ on the basis of
numerical work on the mobilities of touching spheres
\cite{COO,NIR}. Cichocki and Felderhof \cite{FE2} evaluated
$c_{0}=1.454$ (Eq.~(\ref{c0})) by numerically summing their series
expansions of the hydrodynamic interactions while keeping track of
a logarithmic singularity at close separations. Here we reanalyze
$h(x)$ and go well beyond previous computations \cite{FE2,JEF} in
order to gain more insight into the nature of the singularity and
to calculate the residual $R$.

We assume the interaction $U(x)$ is of short range so we focus
only on $h(x)$ for $x-2\lesssim 1$. First, we get an expression
for $A_{11}^{tt}+A_{12}^{tt}$ as an infinite sum from the results
of Stimson and Jeffery\cite{STI} who expressed the hydrodynamic problem in terms of bispherical coordinates. (Note that there is an error in their
paper as pointed out in, for example, Ref.~\onlinecite{BRE} in
which one may find a similar expression for
$A_{11}^{tt}-A_{12}^{tt}$ in case one needs $A_{11}^{tt}$ and
$A_{12}^{tt}$ separately). Calculating $B_{11}^{tt}$ and
$B_{12}^{tt}$ is more involved. We use the numerical scheme by
O'Neill and Majumdar\cite{ONE} which is similar to that of Goldman
et al \cite{GOL}. (Note that there are a few typographical errors
in Ref.~\onlinecite{ONE}. In their Eq.~(3.9) $d$ should be
$d_{1}$, the expression for $v$ in Eq.~(4.1) should have a minus
sign, $\xi$, $\phi$ and $\psi$ in Eqs.~(4.3)-(4.5) should be
replaced by $c\xi$, $c\phi$ and $c\psi$ respectively, and
$\sinh^{2}\lvert\beta\rvert$ in Eq.~(5.10) should be
$\sinh^{3}\lvert\beta\rvert$. Also, to obtain $D_{2}(A_{n},B_{n})$
(Eq.~(3.29)) from $D_{1}(A_{n},B_{n})$ (Eq.~(3.28)) the signs of
$\delta_{n-1}$, $\delta_{n}$ and $\delta_{n+1}$ should be reversed
as well (we only checked the case of spheres of equal size). Their
Table I is correct, however, for spheres of equal size, apart from
the value for $g_{12}(1,0.1)$ which should read -0.1017 instead of
-1.1017).

In order to investigate the regime of lubrication for a pair of
spheres moving under the action of applied forces normal to their
line of centers, we performed the numerical analysis down to
$r/a-2=10^{-10}$ which implies two million terms in the series
expansions are needed. We attempted to speed up the iteration by
adapting the recurrence relationships introduced more recently by
O'Neill and Bhatt \cite{ON2} for a sphere moving near a wall to
the case of two spheres. However, this did not turn out to be
useful as it is for the wall configuration \cite{CHA}. One way of
circumventing series expansions could be to elaborate on the trial
functions initially used by Fixman in his variational theorem for
the mobility matrix \cite{FIX} but we did not investigate this.

\begin{figure}[htb!]
 \begin{center}
%
%
\begin{psfrags}%
\psfragscanon%
%
\psfrag{s05}[][]{\color[rgb]{0,0,0}\setlength{\tabcolsep}{0pt}\begin{tabular}{c}$s$\end{tabular}}%
\psfrag{s06}[][]{\color[rgb]{0,0,0}\setlength{\tabcolsep}{0pt}\begin{tabular}{c}$h$\end{tabular}}%
%
\psfrag{x01}[t][t]{0}%
\psfrag{x02}[t][t]{0.02}%
\psfrag{x03}[t][t]{0.04}%
\psfrag{x04}[t][t]{0.06}%
\psfrag{x05}[t][t]{0.08}%
\psfrag{x06}[t][t]{0.1}%
\psfrag{x07}[t][t]{0.12}%
%
\psfrag{v01}[r][r]{1.31}%
\psfrag{v02}[r][r]{1.32}%
\psfrag{v03}[r][r]{1.33}%
\psfrag{v04}[r][r]{1.34}%
\psfrag{v05}[r][r]{1.35}%
\psfrag{v06}[r][r]{1.36}%
\psfrag{v07}[r][r]{1.37}%
%
\resizebox{8cm}{!}{\includegraphics{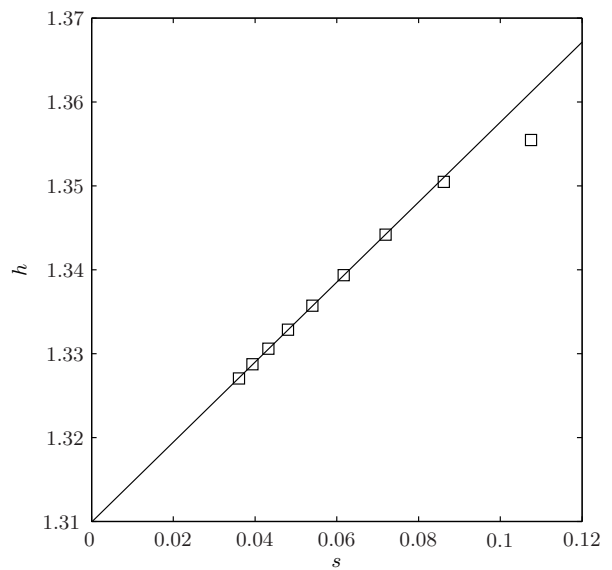}}%
\end{psfrags}%
%

   \caption{The hydrodynamic function $h$ plotted in terms of
the variable $s=-1/(\ln(x-2)-4.694)$. Squares denote results from
the numerical analysis to the accuracy as explained in the text.
The straight line signifies the function $h=0.47666s+1.30993$.\label{fig1}}
 \end{center}
\end{figure}

Goldman et al \cite{GOL} were the first to give a comprehensive
analysis of the mobility of a pair of identical spheres of
arbitrary orientation. They numerically solved the Stokes and
continuity equations using expansions in terms of bipolar
coordinates to high order. For moving spheres whose line of
centers is perpendicular to the applied force, the force consists
not only of a term arising from pure translation but also a term
stemming from pure rotation of the spheres. The latter involves a
torque on one sphere diverging as \cite{GOL}
\begin{equation}
T_{r}\sim \frac{3\ln (x-2)}{160\pi\eta\Omega a^{3}}\label{Tr}
\end{equation}
at very small separations where $\Omega$ is its angular velocity.
Eq. (\ref{Tr}) was derived by extending the nontrivial lubrication
theory of Ref.~\onlinecite{GO2} in which inner and outer regions
have to be matched. Eq. (\ref{Tr}) ultimately leads to the
following analytical expression for $h(x)$ valid at small
separations
\begin{equation}
h(x)=h(2)-\frac{0.47666}{\ln(x-2)+c_{2}}+O(x-2).\label{hx2}
\end{equation}
The coefficient 0.47666 is computed from the numerical tables
presented in Ref.~\onlinecite{GOL}. We have added a constant
$c_{2}$ to the logarithm because we expect the next higher order term in
Eq.~(\ref{Tr}) to be a constant judging by the earlier
analysis of the sphere-wall problem \cite{GO2}. In Fig.~\ref{fig1} we have
fitted Eq.~(\ref{hx2}) to the numerical results discussed above,
letting $h(2)$ and $c_{2}$ be adjustable. The intercept
$h(2)=1.30993$ turns out to be close to the value 1.312 quoted
above for touching spheres which lends credence to the validity of the asymptotic expression that we propose. Moreover, the resulting coefficient $c_{2}=-4.694$ and the concomitant shift in Eq.~(\ref{Tr}) are consistent with the numerical values of the torque $T_{r}$ at small separations as presented in table 3 of Ref.~\onlinecite{GOL}.

\begin{figure}[htb!]
 \begin{center}
%
%
\begin{psfrags}%
\psfragscanon%
%
\psfrag{s05}[][]{\color[rgb]{0,0,0}\setlength{\tabcolsep}{0pt}\begin{tabular}{c}$x$\end{tabular}}%
\psfrag{s06}[][]{\color[rgb]{0,0,0}\setlength{\tabcolsep}{0pt}\begin{tabular}{c}$h$\end{tabular}}%
\psfrag{s07}[][]{\color[rgb]{0,0,0}\setlength{\tabcolsep}{0pt}\begin{tabular}{c}1.31\end{tabular}}%
\psfrag{s08}[][]{\color[rgb]{0,0,0}\setlength{\tabcolsep}{0pt}\begin{tabular}{c}1.33\end{tabular}}%
\psfrag{s09}[][]{\color[rgb]{0,0,0}\setlength{\tabcolsep}{0pt}\begin{tabular}{c}1.35\end{tabular}}%
\psfrag{s10}[][]{\color[rgb]{0,0,0}\setlength{\tabcolsep}{0pt}\begin{tabular}{c}1.37\end{tabular}}%
\psfrag{s11}[][]{\color[rgb]{0,0,0}\setlength{\tabcolsep}{0pt}\begin{tabular}{c}2\end{tabular}}%
\psfrag{s12}[][]{\color[rgb]{0,0,0}\setlength{\tabcolsep}{0pt}\begin{tabular}{c}2.05\end{tabular}}%
\psfrag{s13}[][]{\color[rgb]{0,0,0}\setlength{\tabcolsep}{0pt}\begin{tabular}{c}2.1\end{tabular}}%
\psfrag{s14}[][]{\color[rgb]{0,0,0}\setlength{\tabcolsep}{0pt}\begin{tabular}{c}$x$\end{tabular}}%
\psfrag{s15}[][]{\color[rgb]{0,0,0}\setlength{\tabcolsep}{0pt}\begin{tabular}{c}$h$\end{tabular}}%
%
\psfrag{x01}[t][t]{2}%
\psfrag{x02}[t][t]{2.2}%
\psfrag{x03}[t][t]{2.4}%
\psfrag{x04}[t][t]{2.6}%
\psfrag{x05}[t][t]{2.8}%
\psfrag{x06}[t][t]{3}%
%
\psfrag{v01}[r][r]{0.9}%
\psfrag{v02}[r][r]{0.95}%
\psfrag{v03}[r][r]{1}%
\psfrag{v04}[r][r]{1.05}%
\psfrag{v05}[r][r]{1.1}%
\psfrag{v06}[r][r]{1.15}%
\psfrag{v07}[r][r]{1.2}%
\psfrag{v08}[r][r]{1.25}%
\psfrag{v09}[r][r]{1.3}%
\psfrag{v10}[r][r]{1.35}%
\psfrag{v11}[r][r]{1.4}%
%
\resizebox{8cm}{!}{\includegraphics{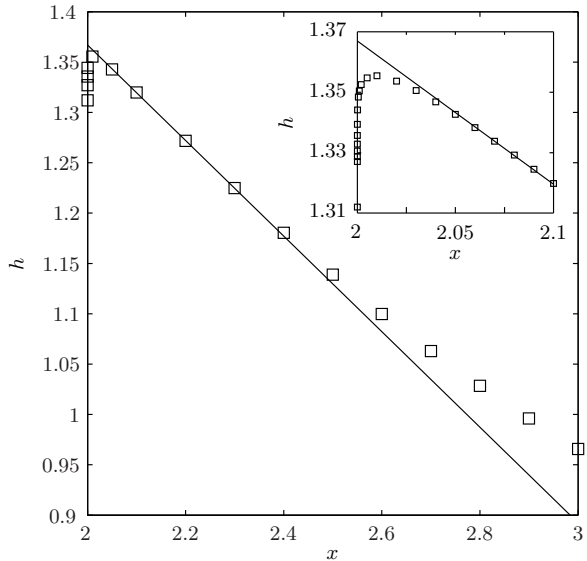}}%
\end{psfrags}%
%

   \caption{The hydrodynamic function $h(x)$ as a function of
the dimensionless separation $x\equiv r/a$ between the centers of
two spheres. The straight line signifies $h_{0}(x)$ given by Eq.~(\ref{inth}).\label{fig2}}
 \end{center}
\end{figure}

Next, we derive an expression for the residual term given by
Eq.~(\ref{R}). First, we propose an initial estimate $h_{0}(x)$ for $h(x)$. We have plotted the numerical values of $h(x)$ as a function of $x$ in Fig.~\ref{fig2}. As a result of the lubrication regime, $h$ has a maximum as displayed in the inset. However, $h(x)$ is only a strongly varying function for $x<2.04$. We therefore simply force a linear fit to the data for $h$ at $x=2.1$, 2.2 and 2.3
\begin{equation}
h_{0}(x)\approx 1.3670-0.4745(x-2).\label{inth}
\end{equation}
We then insert this estimate into Eq.~(\ref{R}) and add a
correction term so as to derive an expression for $R$ accurate
enough for our purposes.
\begin{align}
R &\approx
-0.147\left(\frac{B_{2}}{B_{2}^{HS}}-1\right)\nonumber \\ & +0.4745\int_{2}^{\infty}
\diff{x}x^{2}\left(x-2\right)\left(1-\eu^{-U(x)}\right)
& \nonumber \\ &+9\times10^{-4}\left(1-\eu^{-U(2)}\right).\label{Rap}
\end{align}
The first term on the right comes from the fact that the linear
interpolation gives $h_{0}(2)=1.3670$ whereas the real value is
$h(2)=1.312$. Since the interaction usually does not change
appreciably for $2<x<2.04$, it is straightforward to write an
estimate for the error---the third term---owing to the deviation
of Eq.~(\ref{inth}) from the exact function $h(x)$ (see inset
Fig.~\ref{fig2}). In our case the error term turns out to be an order of
magnitude smaller than the first two terms.

\subsection*{E. Determination of $\lambda_{C}$}

It is clear from Eq.~(\ref{R}) that $R$ would vanish if the actual
interaction were a pure AHS potential. If we then insert
Eq.~(\ref{B2bax}) into Eq.~(\ref{aplc2}), we obtain \cite{FE1}
\begin{equation}
\lambda_{C}= c_{0}-\frac{2c_{1}}{3\tau}.\label{lcbax}
\end{equation}
Inspection of the various terms in Eq.~(\ref{Rap}) reveals that
$R$ is often much smaller than unity when the
interaction is given by Eq.~(\ref{pot}). Hence, a possibly
convenient approximation to the coefficient $\lambda_{C}$ is from
Eq.~(\ref{tb2})
\begin{equation}
\lambda_{C}= c_{0}+c_{1}J=c_{0}+\frac{8c_{1}}{3}
\left(\frac{B_{2}}{B_{2}^{HS}}-1\right) \label{lcappr}
\end{equation}
where $J$ may be evaluated numerically or approximately with the
help of Eqs.~(\ref{vir3})-(\ref{J2}).

The full expression for the dynamical coefficient is written as
\begin{equation}
\lambda_{C}=c_{0}+c_{1}J+R,\label{lct}
\end{equation}
using Eqs.~(\ref{B2bax}) and (\ref{aplc2}). Now $R$ from
Eq.~(\ref{Rap}) is reexpressed as
\begin{equation}
R\approx -0.055J+0.4745K
-9\times10^{-4}\left(\eu^{U_{A}-\xi}-1\right).\label{R2}
\end{equation}
in view of Eqs.~(\ref{pot}) and (\ref{debye}). Here we have
introduced the function $K$ for which we derive a convenient
approximation.
\begin{align}
K&\equiv\int_{2}^{\infty}\diff{x}x^{2}\left(x-2\right)
\left(1-\eu^{-U_{T}(x)}\right)\nonumber \\ & \equiv K_{1}-\left(
\eu^{U_{A}}-1\right) K_{2}\label{K},
\end{align}
where
\begin{equation}
K_{1}\equiv\int_{2}^{\infty}\diff{x}x^{2}\left(x-2\right)\left(
1-\eu^{-U_{DH}(x)}\right)\label{K1}
\end{equation}
and
\begin{equation}
K_{2}\equiv\int_{2}^{2+\delta}\diff{x}x^{2}\left(x-2\right)
\eu^{-U_{DH}(x)}\label{K2}.
\end{equation}
In the same spirit as in Ref.~\onlinecite{PR1}, we approximate
$x\left( 1-\eu^{-U_{DH}(x)}\right)\approx 2\xi \eu^{-\mu
(x-2)}-2\alpha\xi^{2}\eu^{-2\mu(x-2)}$, with
$\alpha=\frac{\eu^{-\xi}-(1-\xi)}{\xi^{2}}$. We then have
\begin{equation}
K_{1}\approx \frac{\xi\left(
\mu+1\right)\left(4-\alpha\xi\right)}{\mu^{3}}\label{in1},
\end{equation}
where we have neglected the small term $\alpha\xi^{2}/2\mu^{3}$.
In the case of lysozyme at pH 4.5, the deviation of
Eq.~(\ref{in1}) from the exact result is smaller than about $3\%$
for $I\geq0.05$ M and smaller than about $1\%$ for $I\geq0.3$ M.
For the second integral we use the trapezoid approximation
$\int_{2}^{2+\delta }\diff{x}g(x)\approx\frac{1}{2}\delta\left[
g(2)+g(2+\delta)\right]$ ($\delta\ll 1$) and we neglect a factor
$(1+\delta/2)^{2}$
\begin{equation}
K_{2}\approx 2\delta^{2}\exp\left[{-\frac{\xi
\eu^{-\mu\delta}}{1+\delta/2}}\right]\label{in2}.
\end{equation}
For lysozyme at pH 4.5 with $\delta=0.079$ (see below), this
approximation deviates less than about $5\%$ from the exact value
for $I\geq0.05$ M and less than about $3\%$ for $I\geq0.2$ M.

\section*{III. Comparison with experiment}

\begin{figure}[htb!]
 \begin{center}
%
%
\begin{psfrags}%
\psfragscanon%
%
\psfrag{s05}[][]{\color[rgb]{0,0,0}\setlength{\tabcolsep}{0pt}\begin{tabular}{c}$I$ (M)\end{tabular}}%
\psfrag{s06}[][]{\color[rgb]{0,0,0}\setlength{\tabcolsep}{0pt}\begin{tabular}{c}$\lambda_{C}$\end{tabular}}%
%
\psfrag{x01}[t][t]{0}%
\psfrag{x02}[t][t]{0.5}%
\psfrag{x03}[t][t]{1}%
\psfrag{x04}[t][t]{1.5}%
\psfrag{x05}[t][t]{2}%
%
\psfrag{v01}[r][r]{-30}%
\psfrag{v02}[r][r]{-20}%
\psfrag{v03}[r][r]{-10}%
\psfrag{v04}[r][r]{0}%
\psfrag{v05}[r][r]{10}%
\psfrag{v06}[r][r]{20}%
\psfrag{v07}[r][r]{30}%
%
\resizebox{8cm}{!}{\includegraphics{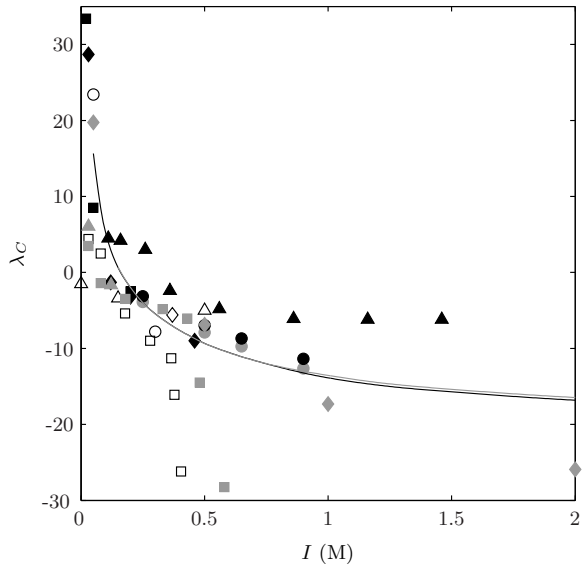}}%
\end{psfrags}%
%

   \caption{Experimental data and theoretical predictions of
$\lambda_{C}$ for lysozyme as a function of the ionic strength $I$
at a pH of about 4.5. Data: black squares: Nystr\"om et al.
\cite{NYS}, pH 4.0, 25 $\operatorname{{}^{\circ}{\rm C}}$; grey
squares: Mirarefi et al. \cite{MIR}, pH 4.6; white squares:
Mirarefi et al. \cite{MIR}, pH 4.6; black diamonds: Muschol et al.
\cite{MUS}, pH 4.7, 20 $\operatorname{{}^{\circ}{\rm C}}$; grey
diamonds: Zhang et al. \cite{ZHA}, pH 4.5, 20
$\operatorname{{}^{\circ}{\rm C}}$; white diamonds: Skouri et al.
\cite{SKO}, pH 4.6, 20 $\operatorname{{}^{\circ}{\rm C}}$; black
triangles: Eberstein et al. \cite{EBE}, pH 4.2, 20
$\operatorname{{}^{\circ}{\rm C}}$; grey triangles: Leggio et al.
\cite{LEG}, pH 4.75, 25 $\operatorname{{}^{\circ}{\rm C}}$; white
triangles: Price et al. \cite{PRI}, pH 4.6, 25
$\operatorname{{}^{\circ}{\rm C}}$; black circles: Annunziata et
al. \cite{ANN}, pH 4.5, 25 $\operatorname{{}^{\circ}{\rm C}}$ ;
grey circles: Annunziata et al. \cite{ANN}, pH 4.5, 25
$\operatorname{{}^{\circ}{\rm C}}$; white circles: Retailleau et
al. \cite{RET}, pH 4.0. In all cases, the supporting electrolyte
is NaCl, often with a small amount of Na acetate added. The grey
line denotes the theoretical curve setting $R\equiv 0$ i.e.
Eq.~(\ref{lcbax}) with $\tau$ given by Eq.~(\ref{tb2}), and the
black line is the curve given by Eq.~(\ref{lct}). The functions
$J$ and $K$ have been approximated as outlined in the text.\label{fig3}}
 \end{center}
\end{figure}

We compare our predictions of $\lambda_{C}$ as a function of the
ionic strength $I$ with experimental results for lysozyme at room
temperature and at a pH of about 4.5. The added salt is NaCl and
in most cases a small amount of Na acetate has been added as buffer. The reason for choosing lysozyme under these conditions is that we have
previously evaluated the range and strength of the short-range
attraction \cite{PR1} and a lot of experimental data on the
collective diffusion coefficient are available in the literature
(see Fig.~\ref{fig3}).

Lysozyme has a moderate aspect ratio of about 1.5 and we
approximate it by a sphere of radius $a=1.7$ nm \cite{MIK}. The
dimensionless parameter $\mu$ is then given by $\mu=5.58\sqrt{I}$,
where the ionic strength $I$ is given in M, and
$\xi=0.209(\overline{Z}/(1+\mu))^{2}$. Here we follow our
discussion in Ref.~\onlinecite{PR1} and use the adjusted charge on
the lysozyme sphere $\overline{Z}=Z_{eff}-1$ instead of the
effective charge $Z_{eff}$. Values of $Z$, $Z_{eff}$ and
$\overline{Z}$ as a function of ionic strength can be found in
Table I as well as the corresponding quantities $\mu$ and $\xi$.
For the range $\delta$ and strength $U_{A}$ of the attraction we
use $\delta=0.079$ and $U_{A}=3.70$ which were computed on the
basis of a wide variety of data on the second virial coefficient
\cite{PR1}.

\begin{table*}[htb!]

\begin{center}

\begin{tabular}{lrrrlrrlrrr}

\hline

 & & & & & & & & & $\lambda_{C}$ \hspace{1mm} & \\

\raisebox{1.5ex}[0pt]{$I$ (M)} \hspace{0mm} &
\raisebox{1.5ex}[0pt]{$Z$} \hspace{0.5mm} & \hspace{3mm}
\raisebox{1.5ex}[0pt]{$Z_{eff}$} & \hspace{3mm}
\raisebox{1.5ex}[0pt]{$\overline{Z}$} \hspace{0.2mm} &
\hspace{5mm} \raisebox{1.5ex}[0pt]{$\xi$} \hspace{1mm} &
\hspace{5mm} \raisebox{1.5ex}[0pt]{$\mu$} \hspace{1mm} &
\hspace{4mm} \raisebox{1.5ex}[0pt]{$R$} \hspace{3mm} &
\hspace{6mm} \raisebox{1.5ex}[0pt]{$\tau$} \hspace{3mm} &
\hspace{4mm} via $\tau$ & \hspace{3mm} direct
& \hspace{3mm} exact \\

\hline \hline

$0.05$ & $9.5$ & $8.8$ \hspace{0.5mm} & $7.8$ & \hspace{2mm} $2.52$ & $1.25$ & 4.267 & & & 16.63 & $15.66$ \\

$0.10$ & $9.8$ & $9.2$ \hspace{0.5mm} & $8.2$ & \hspace{2mm} $1.84$ & $1.76$ & 1.382 & & & 5.30 & $5.14$ \\

$0.15$ & $10.0$ & $9.4$ \hspace{0.5mm} & $8.4$ & \hspace{2mm} $1.48$ & $2.16$ & $0.744$ & \hspace{2mm} $0.742$ & $-0.06$ & $0.68$ & $0.65$ \\

$0.20$ & $10.1$ & $9.6$ \hspace{0.5mm} & $8.6$ & \hspace{2mm} $1.27$ & $2.50$ & $0.514$ & \hspace{2mm} $0.286$ & $-2.48$ & $-1.97$ & $-1.96$ \\

$0.25$ & $10.2$ & $9.7$ \hspace{0.5mm} & $8.7$ & \hspace{2mm} $1.10$ & $2.79$ & $0.408$ & \hspace{2mm} $0.193$ & $-4.38$ & $-3.97$ & $-3.95$ \\

$0.30$ & $10.2$ & $9.8$ \hspace{0.5mm} & $8.8$ & \hspace{2mm} $0.984$ & $3.06$ & $0.354$ & \hspace{2mm} $0.155$ & $-5.81$ & $-5.46$ & $-5.45$ \\

$0.45$ & $10.3$ & $10.0$ \hspace{0.5mm} & $9.0$ & \hspace{2mm} $0.752$ & $3.74$ & $0.302$ & \hspace{2mm} $0.109$ & $-8.82$ & $-8.53$ & $-8.51$ \\

$1.0$ & $10.4$ & $10.2$ \hspace{0.5mm} & $9.2$ & \hspace{2mm} $0.409$ & $5.58$ & $0.323$ & \hspace{2mm} $0.0734$ & $-13.87$ & $-13.55$ & $-13.58$ \\

$1.5$ & $10.4$ & $10.3$ \hspace{0.5mm} & $9.3$ & \hspace{2mm} $0.295$ & $6.83$ & $0.349$ & \hspace{2mm} $0.0655$ & $-15.72$ & $-15.38$ & $-15.43$ \\

$2.0$ & $10.4$ & $10.3$ \hspace{0.5mm} & $9.3$ & \hspace{2mm} $0.229$ & $7.89$ & $0.367$ & \hspace{2mm} $0.0616$ & $-16.82$ & $-16.46$ & $-16.52$ \\

\hline

\end{tabular}

\end{center}

\caption{Values of the actual charge $Z$ of hen-egg-white lysozyme
(from Ref.~\onlinecite{KUE}), the effective charge $Z_{eff}$ (see
Eq.~(\ref{Zeff})), the lowered effective charge
$\overline{Z}=Z_{eff}-1$, and dimensionless interaction parameters
$\xi$ and $\mu$ as a function of the ionic strength $I$. The pH
equals $4.5$ and $\xi$ has been calculated using the lowered
effective charge $\overline{Z}$. $R$ has been calculated from
Eq.~(\ref{R2}), $\tau$ from Eq.~(\ref{tb2}), $\lambda_{C}$ (via
$\tau$) from Eq.~(\ref{lcbax}) and $\lambda_{C}$ (direct) from
Eq.~(\ref{lct}). In all cases approximations for $J$ and $K$ given
by Eqs.~(\ref{vir3})-(\ref{J2}) and (\ref{K}), (\ref{in1}),
(\ref{in2}) were used. The computation of the numerically exact
$\lambda_{C}$ is explained in the text.}

\end{table*}

We next employ three methods to predict $\lambda_{C}$
theoretically. In the first, we compute $\tau$ by equating the
respective second virial coefficients of Section II.B (see
Eqs.~(\ref{vir3}) and (\ref{tb2})). We then calculate
$\lambda_{C}$ from Eq.~(\ref{lcbax}) using $c_{0}=1.454$ and
$c_{1}=1.688$. In the second method we use Eq.~(\ref{lct}) to
determine $\lambda_{C}$, where $R$ is evaluated with the help of
Eq.~(\ref{R2}). In both cases the approximations for $J$ and $K$
given by Eqs.~(\ref{vir3})-(\ref{J2}) and Eqs.~(\ref{K}),
(\ref{in1}) and (\ref{in2}) were used (see Table I and Fig.~\ref{fig3}).
Note that there are no free parameters so the curves in Fig.~\ref{fig3}
are predictions not fits. For comparison, we also calculate
$\lambda_{C}$ from Eq.~(\ref{lc}) exactly, that is by performing
the integrals in Eqs.~(\ref{lv})-(\ref{la}) numerically with the
help of a highly accurate interpolation formula for $h(x)$ (see
Table I). Finally, in Fig.~\ref{fig3} we have also plotted data of
$\lambda_{C}$ measured by several experimental groups.

\section*{IV. Discussion}

In Section II.E we have outlined two approximate methods to
calculate $\lambda_{C}$. As one can see from Table I, both the
direct method incorporating an approximation for the residual $R$
and the method relying solely on the stickiness $\tau$ via the
second virial yield results that are often close to the exact
numerical computations. The direct method is, of course, somewhat
more accurate. The $\tau$ method breaks down below 0.2 M. Note
that in the important regime $I>0.2$ M pertaining to protein
crystallization, $R$ is much smaller than the absolute magnitude
of $\lambda_{C}$. This may explain why $\lambda_{C}$ is a useful parameter to characterize the onset of crystallization \cite{EBE}.

In Fig.~\ref{fig3}, it is clear that there is a large degree of scatter which may be attributed to the systematic variation in sets of data from the various groups, especially at large ionic strengths ($I>0.4$ M). We do not know what is the cause of this. In one experiment \cite{MIR}, we do observe there is considerable scatter in a plot of the diffusion coefficient versus the protein solubility which might explain the extreme downturn of several data in Fig.~\ref{fig3} at about 0.5 M. Fig.~\ref{fig3} also shows that our predicted curves lie fairly neatly in the midst of the swarm of data. We
emphasize again that we have no adjustable parameters in our
calculations except for a slight downward adjustment of the
effective charge (see also the discussion in
Refs.~\onlinecite{PR1} and \onlinecite{PR2}). The model is thus not
inconsistent with the experimental data though we will have to await
more experiments under conditions which are better controlled
before one may reach a more definitive conclusion. In a similar vein, it is not possible to claim that the neglect of electrolyte friction assumed
here is entirely warranted.

In summary, we have approximated proteins by spherical particles
interacting by a hard-core and electrostatic repulsion together
with a short-range attraction. An analysis of the two-particle
statistics and hydrodynamics leads to a reasonable prediction of
the ionic-strength dependence of the linear coefficient
$\lambda_{C}$. At high ionic strengths, when $B_{2}$ is negative,
the residual $R$ is relatively small so there is then an
interesting direct relationship between $\lambda_{C}$ and $B_{2}$
(Eq.~(\ref{lcappr})) which could be tested experimentally.

\end{document}